\documentclass[%
superscriptaddress,
twocolumn,
amsmath,amssymb,
 aps,
prb,
floatfix,
]{revtex4-2}

\usepackage{graphicx}
\usepackage[colorlinks]{hyperref}
\hypersetup{%
	plainpages=true,
	breaklinks=true,
	hypertexnames=false,
	pageanchor=true,
	colorlinks=true,
	linkcolor={blue},
	citecolor={red},
	urlcolor={blue},
	anchorcolor={black}
}

\usepackage[dvipsnames]{xcolor}
\usepackage[normalem]{ulem}

\begin{document}
\title{
Exact solution for quantum strong long-range models via a generalized Hubbard-Stratonovich transformation
}

\author{Juan Román-Roche}
\affiliation {Instituto de Nanociencia y Materiales de Aragón (INMA), CSIC-Universidad de Zaragoza, Zaragoza 50009, Spain}
\affiliation{Departamento de Física de la Materia Condensada, Universidad de Zaragoza, Zaragoza 50009, Spain}

\author{Víctor Herráiz-López}
\affiliation {Instituto de Nanociencia y Materiales de Aragón (INMA), CSIC-Universidad de Zaragoza, Zaragoza 50009, Spain}
\affiliation{Departamento de Física de la Materia Condensada, Universidad de Zaragoza, Zaragoza 50009, Spain}

\author{David Zueco}
\affiliation {Instituto de Nanociencia y Materiales de Aragón (INMA), CSIC-Universidad de Zaragoza, Zaragoza 50009, Spain}
\affiliation{Departamento de Física de la Materia Condensada, Universidad de Zaragoza, Zaragoza 50009, Spain}
  
\date{\today}

\begin{abstract}
We present an exact analytical solution for quantum \emph{strong} long-range models in the canonical ensemble by extending the classical solution proposed in [Campa \emph{et al.}, J. Phys. A 36, 6897 (2003)]. Specifically, we utilize the equivalence between generalized Dicke models and interacting quantum models as a generalization of the Hubbard-Stratonovich transformation. To demonstrate  our method, we apply it to the Ising chain in transverse field and discuss its potential application to other models, such as the Fermi-Hubbard model, combined short- and long-range models and models with antiferromagnetic  interactions.  Our findings indicate that the critical behaviour of a model is independent of the range of interactions, within the strong long-range regime, and the dimensionality of the model. Moreover, we show that the order parameter expression is equivalent to that provided by mean-field theory, thus confirming the exactness of the latter.  Finally, we examine the algebraic decay of correlations and characterize its dependence on the range of interactions in the full phase diagram.\end{abstract}

\maketitle

\section{Introduction}
\label{intro}

Long-range systems are those in which two-body interactions decay as a power-law at large distances. They are ubiquitous in nature, with some examples given by dipolar, Coulomb or Wan-der-Walls interactions. Recent experimental advances in atomic, molecular and optical systems have lead to a resurgence of interest in  long-range models \cite{Britton2012, knap2013probing, monroe2021programmable, browaeys2020manybody}. In these experiments, the effective interactions between spins are often long-ranged and tunable, renewing the need for a comprehensive  understanding of long-range systems.
Although less studied than their short-ranged counterparts, there are already some rigorous and numerical results available \cite{mukamel2008statistical, campa2009statistical, fey2016critical, defenu2023longrange, defenu2023outofequilibrium}.  
Some equilibrium and dynamical properties have been discussed in comparison with short-range systems. Notable examples are the existence (or absence) of an area law of entanglement \cite{koffel2012entanglement, kuwahara2020area, vodola2014kitaev, ares2018entanglement}, the algebraic decay of two-point correlators out of criticality \cite{vodola2015longrange, vanderstraeten2018quasiparticles, francica2022correlations}, the spreading of correlations \cite{schneider2022spreading}, the existence of Majorana modes \cite{jager2020edge} and topological properties \cite{viyuela2016topological}. 

In these examples, the phenomenology can be understood within  a (sub)classification in terms of the range of interactions they exhibit. 
To fix notation and ideas, let us introduce this classification with the models considered in this paper: quantum long-range models in an $N$-site lattice with a coupling of the form
\begin{equation}
 	\mathcal H_{\rm c}
  =
  -\sum_{ij}^N J_{ij} \mathcal C_i \mathcal C_j\,,	\label{eq:longrangecoupling}
\end{equation}
where $\mathcal C_i$ is a local hermitian operator acting on site $i$. We consider models with power-law decaying interactions $J_{ij} = \Gamma \tilde J(\mathbf r_{ij}) / \tilde N$,
\begin{equation}
\label{eq:Jalpha}
	\tilde J(\mathbf r_{ij})= \begin{cases}
	b & \text { if } \quad \mathbf r_{ij}=0 \\
	|\mathbf r_{i j}|^{-\alpha} & \text { otherwise}
	\end{cases}
\end{equation}
and periodic boundary conditions (PBC). 
\begin{figure}
\centering
\includegraphics[width=\columnwidth]{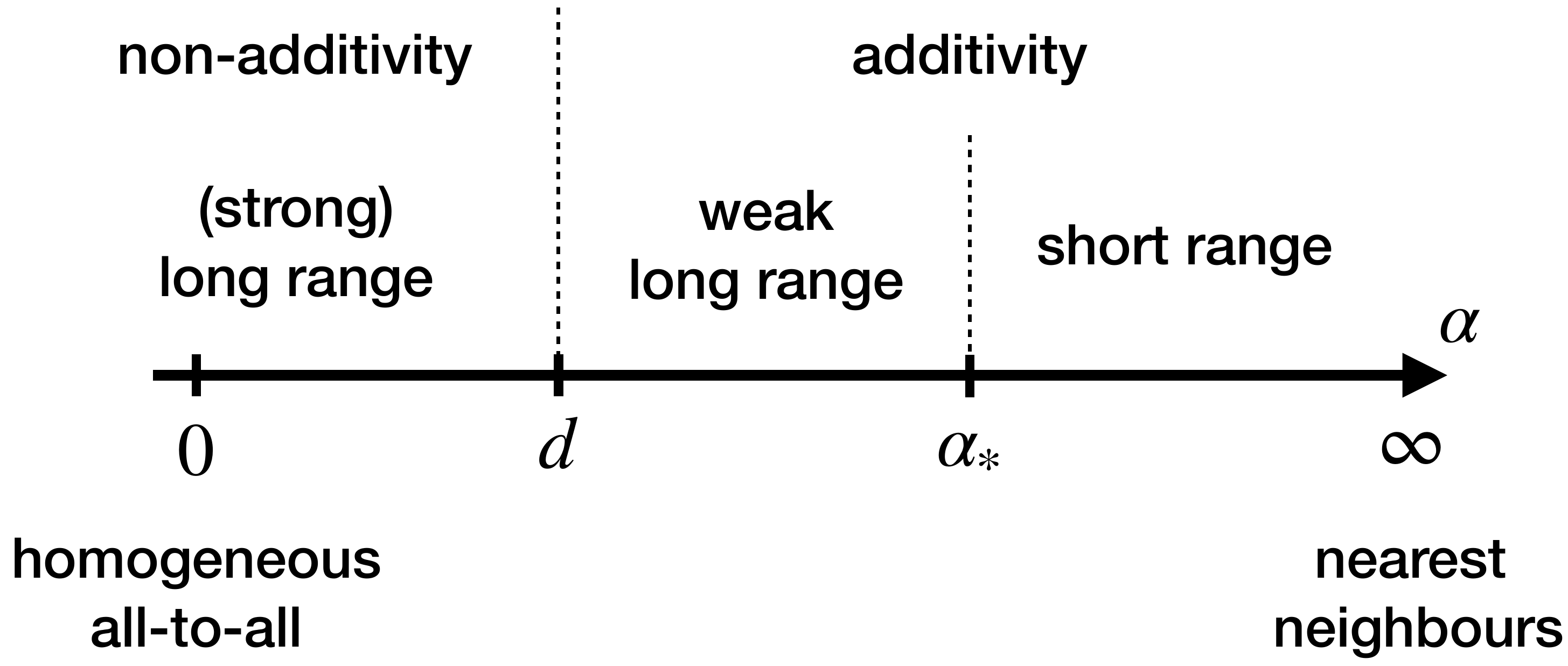}
\caption{Classification of long-range models following reviews \cite{mukamel2008statistical} and \cite{defenu2023longrange} valid for both  the classical and quantum models.  This work presents a solution for the quantum strong long-range regime.} 
\label{fig:longcla}
\end{figure}
The distance between sites $\mathbf r_{ij}$ is then given by the nearest image convention. Through this work we will focus on the case of attractive or ferromagnetic interaction, so the interaction strength is $\Gamma > 0$, although the extension to antiferromagnetic or repulsive models will be discussed.  $b$ is a parameter that can be tuned to shift the spectrum of $J$. The decay rate, $\alpha$, sets the range of the interactions.  For $\alpha < d$, where $d$ is the dimensionality of the lattice, the interactions decay slowly enough that the sum in the coupling term \eqref{eq:longrangecoupling} depends superlinearly on $N$, breaking the extensivity of the model (See App. \ref{extensivity}.). Kac's renormalization factor $1/\tilde N$ restores extensivity, ensuring a well-defined thermodynamic limit. Here $\tilde N = \sum_i \tilde J_{ij}$, note that PBC make the model translation invariant and thus $\sum_i \tilde J_{ij}$ is independent of $j$. Regardless, the model remains non-additive in this regime. Non-additivity brings about particular statistical and dynamical phenomena that differ from the commonly studied short-range models, such as ensemble inequivalence, negative specific heat and quasistationary states \cite{defenu2023longrange}. Accordingly, the regime $\alpha < d$ is identified as \emph{(strong) long-range}. In the regime $\alpha > d$, the model is naturally extensive, and Kac's renomarlization factor amounts to a rescaling of the interaction strength. Within the regime $\alpha > d$ two further subregimes can be identified: for $\alpha > \alpha_*$  the critical exponents of the model match those of the nearest-neighbours model ($\alpha \to \infty$), this is the \emph{short-range regime}; for $d < \alpha < \alpha_*$ the model presents critical exponents that differ from the short-range ones, the effects of long-range interactions are felt but the model is additive, this is the \emph{weak long-range regime} \cite{mukamel2008statistical, defenu2023longrange}.
For convenience, we summarize this classification in Fig. \ref{fig:longcla}.  
\emph{This work deals with the strong long-range regime}.

Strong long-range models are commonly disregarded in many analytical and numerical studies on the grounds of the ill-defined thermodynamic limit brought about by the non-extensivity. Kac's rescaling eliminates this barrier, making their study possible. For quantum models,  seminal numerical studies solving the transverse-field Ising model in the strong long-range regime are found in Refs. \cite{koziol2021quantumcritical, gonzalezlazo2021finitetemperature}.   They confirm that in this regime the model is within  the mean-field universality class.  
This is in agreement with the claim that mean field is exact for  quantum spin models in the strong long-range regime \cite{mori2012equilibrium} that generalizes similar findings in classical systems \cite{campa2000canonical,mori2011instability,mori2010analysis,mori2012microcanonical}.  These works are crucial for the rigorous understanding of the physics of long-range systems.  On the one hand, they provide an exact way to solve them, on the other hand, they provide a starting point for approximations that tackle the weak long-range regime.

This work provides a recipe to analytically solve, in the canonical ensemble, quantum strong long-range models.  Therefore, it complements the work of Mori \cite{mori2012equilibrium} and confirms that \emph{in the strong long-range regime  mean field is exact}.  Besides, it  extends the work of Campa and coworkers for classical strong long-range models to the quantum case \cite{campa2003canonical}.
Our work introduces a generalized Hubbard-Stratonovich transformation (HST) and provides a closed expression for the free energy at any temperature.  
Technically, we show how to use the equivalence between generalized Dicke models and interacting quantum models as a quantum HST. We show that only \emph{strong} long-range models admit this mapping and formulate their canonical solution in terms of the associated Dicke model, which is then tackled following the prescription of Wang and Hioe \cite{wang1973phase, hioe1973phase}. We illustrate the method on the Ising chain in transverse field. We find that the critical behaviour is universal for $\alpha < d$ and any lattice dimensionality. The expression for the magnetization (the order parameter) is shown to be equivalent to the mean-field solution, thus proving the exactness of the latter. Finally, we study the algebraic decay of correlations as a function of the decay rate of interactions $\alpha$.

The rest of the paper is organized as follows. In Section \ref{sec:sketch}, we provide a brief overview of the HST as a tool to solve classical models, which forms the basis for our further development. In Section \ref{sec:generalhs}, we establish the relationship between generalized Dicke and long-range models and introduce the generalized HST. Section \ref{sec:solvedicke} presents the solution for strong long-range models and a discussion of which models can be treated with this method. We perform the calculations for the long-range transverse field Ising model, including the full phase diagram and the decay of two-point correlations in section \ref{sec:ising}. Finally, we conclude the article with some general remarks and relegate more technical details to the Appendices.


\section{Sketch of the solution for classical systems}
\label{sec:sketch}

To warm up, it is convenient to understand how to solve classical strong long range models, mainly following the works of Campa and coworkers \cite{campa2000canonical, campa2003canonical}.  For simplicity consider the Ising model,
\begin{equation}
    \mathcal H_{\rm cl}=
     h\sum_i^N s_i - \sum_{ij}^N J_{ij} s_i s_j \; .
\end{equation}
Here, $s_i$ is a discrete variable. The solution is based on two main observations.
First, diagonalizing the interaction matrix $J = \Lambda D \Lambda^T$,  which allows one to write the coupling term as $\sum_k D_k \left( \sum_i \Lambda_{ik} s_i \right)^2$, where $\{D_k\}$ are the eigenvalues of the interaction matrix. Note that, in this form, the coupling is written as a sum of interaction terms that are quadratic in $\sum_i^N \Lambda_{ik} s_i$. Second, eliminating these quadratic interactions by use of the Hubbard-Stratonovich transformation which is based on the equality,
\begin{equation}
\begin{split}
    Z & = \sum_{s_i}
    e^{- \beta \mathcal H_{cl}}
    \\ 
    &\propto
    \int d u_k
    \sum_{s_i} e^{
    -\beta 
    \big ( 
    h \sum_i s_i
    +  \sum_k u_k^2/D_k 
    - 2 \sum_{ik} \Lambda_{ik}
s_i u_k  \big )
    }
\end{split}
\end{equation}
here, $u_k$ are real auxiliary variables.  This equality for the partition function
 follows from Gaussian integral formulas.

 Notice that we have decoupled the interaction $J_{ij} s_i s_j$, therefore the sum over $s_i$-configurations are trivial.  Finally, the integral over the real variables $u_k$ can be done within the saddle point approximation in the $N$-large limit. This is true if  some conditions are met on the eigenspectrum of the $J_{ij}$ matrix, see below and the discussion in App. \ref{secondordercorrections}. 

 
\section{From the Dicke model to quantum long-range models and back}
\label{sec:generalhs}
\subsection{Effective theory of the Dicke model}

\begin{figure}
    \centering
    \includegraphics[width = 1.0 \columnwidth]{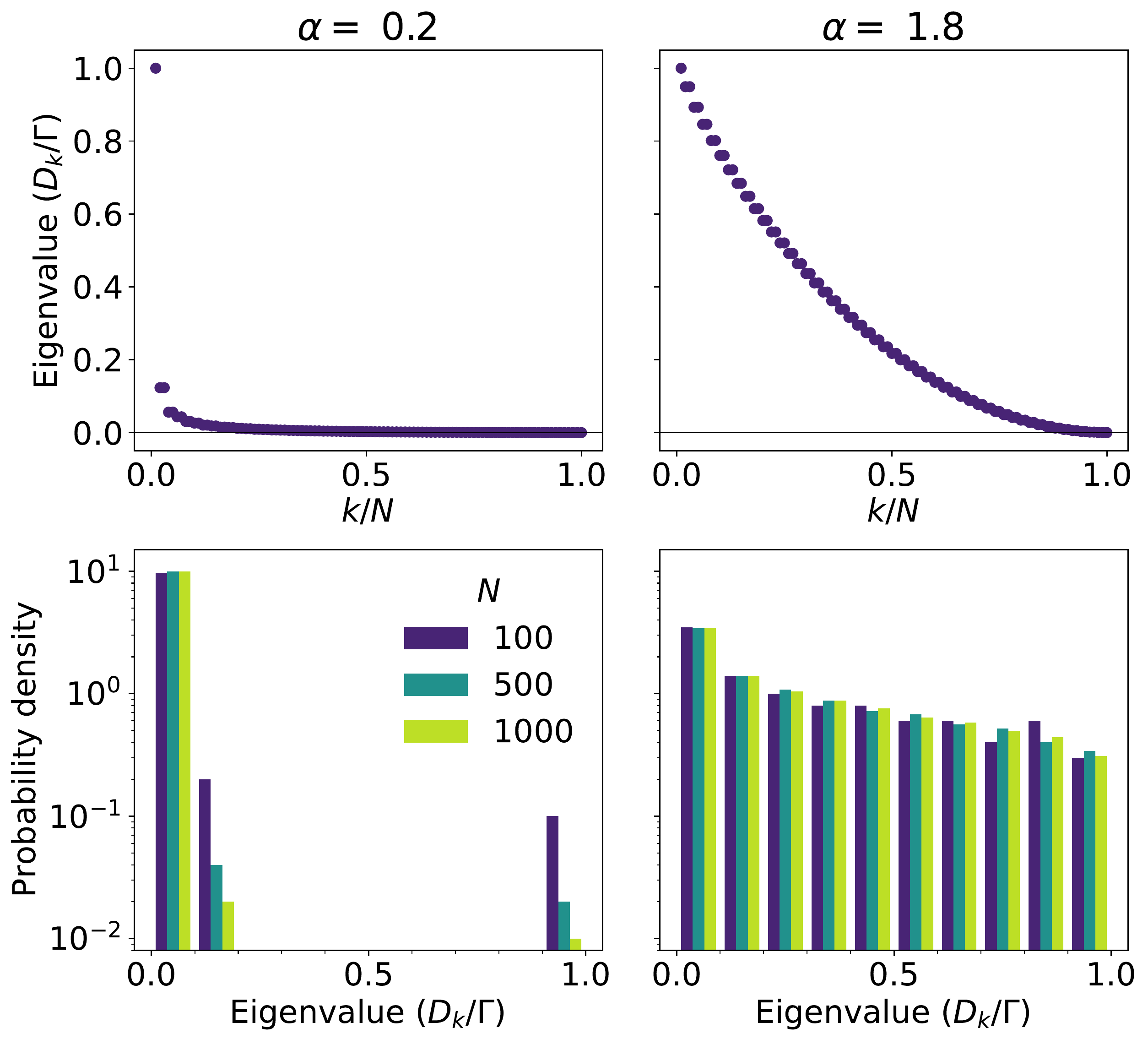}
    \caption{Analysis of the eigenvalues of the coupling matrix $J$ \eqref{eq:longrangecoupling} for $d=1$. Left: For $\alpha=0.2 < 1$, a plot of the eigenvalues for $N=100$ on top and a histogram of the eigenvalues as a function of $N$ on the bottom. Right: Same but for $\alpha=1.8 > 1$.}
    \label{fig:levelstatistics}
\end{figure}

The method described above and utilized in Ref. \cite{campa2003canonical} cannot be straightforwardly applied to quantum models. The application of the HST requires splitting the exponential that constitutes the kernel of the partition function into a product of exponentials, which in the quantum case is prevented by the non-commutativity of the long-range interaction term and other terms in the Hamiltonian. There are ways in which the HST can be applied to solve quantum systems, but it requires a reframing of the partition function in terms of commuting quantities. A field-theory formulation or imaginary-time Trotterization are examples of this. Here we present an alternative which is the closest to the classical formulation. 

Our method utilizes some results from quantum optics in order to draw an equivalence between some quantum long-range models and a cavity QED model. Specifically, we utilize the generalized Dicke model as our starting point to develop this equivalence:
\begin{equation}
	\mathcal H = \sum_{k=0}^{M-1} \omega_k a_k^\dagger a_k + \mathcal H_0 - \sum_{k, i} \left( a_k + a_k^\dagger \right) \frac{\lambda_{i k}}{\sqrt{N}} \mathcal C_i \,.
	\label{eq:dickemodel}
\end{equation}
Here $\mathcal H_0$ is an exactly solvable Hamiltonian of the ``mater'' degrees of freedom and $\mathcal C_i$ is the local hermitian operator that couples site $i$ to the bosonic modes  with $[a_k, a_{k^\prime}^\dagger] = \delta_{k, k^\prime}$, finally $\lambda_{ik}$ are real coupling constants. In two previous publications \cite{romanroche2021photon, romanroche2022effective} we show that the generalized Dicke model, after integrating out the electromagnetic modes, yields an exact effective Hamiltonian description for the matter degrees of freedom alone in the limit $N \to \infty$ (thermodynamic limit). The resulting Hamiltonian,
\begin{equation}
	 \mathcal H_{\rm m}^{\rm eff} = \mathcal H_0-  \sum_{ij}^N \sum_{k=0}^{M-1} \frac{\lambda_{i k} \lambda_{j k}}{N\omega_k} \mathcal C_i \mathcal C_j \,,
 \label{Heff}
\end{equation}
corresponds to a quantum model with interactions given by $(J_{\rm eff})_{ij} = \sum_k^{M-1} \lambda_{i k} \lambda_{j k} / (N \omega_k)$. The mode structure of the cavity determines the resulting effective model. However, it is important to note that the exact mapping between Hamiltonians \eqref{eq:dickemodel} and \eqref{Heff} is limited to the thermodynamic limit, $N \to \infty$, and a number of modes $M$ such that $\lim_{N \to \infty} M/N = 0$. Below, we  demonstrate how we can reverse the effective theory to solve a quantum model.
The first question that arises is which family of quantum models, with interaction given by Eq. \eqref{eq:longrangecoupling}, can be solved this way, \emph{i.e.} which can be cast in the form of Eq. \eqref{Heff}. Below, we show that this is the case for strong long range models, $\alpha < d$, this is the first result of this paper. 


\subsection{Mapping a quantum model to the Dicke model}

If we start from an arbitrary extensive \footnote{meaning that Kac's prescription is used to ensure extensivity if the model is strong-long-ranged} model of the form $\mathcal H_{\rm m} = \mathcal H_0 - \sum_{ij} J_{ij} \mathcal C_i \mathcal C_j$, Cf. Eq. \eqref{eq:longrangecoupling}, the first step is to diagonalize the interaction matrix $J = \Lambda D \Lambda^T$, where $D$ is a diagonal matrix, $D_{kp} \equiv D_{k} \delta_{kp}$. Note that $\Lambda$ is orthogonal because $J$ is symmetric. The matrix elements are then given by 
 \begin{equation}
    J_{ij} = \sum_{k=0}^{N-1} \Lambda_{ik} D_k \Lambda_{jk}
    \label{eq:Jij-decompositon-original}
 \end{equation}
Assuming that $J_{ij} >0$, the smallest eigenvalue of $J$ can always be set to zero by adjusting its diagonal elements, which we denote $b$. Fixing $b \neq 0$ introduces, generally, non-trivial diagonal terms of the form $\Gamma b/\tilde N \mathcal C_i^2$. These can be shown to be negligible in the thermodynamic limit, so the freedom to set $b$ remains (See App. \ref{diagonalterms}.).
For a general interaction matrix the number of non zero eigenvalues, $M$, scales with the size of the matrix, $N$. Conveniently, it can be shown that for a model with power-law decaying interactions and PBC such as the one considered in this work \eqref{eq:longrangecoupling}, the number of non-zero modes in the thermodynamic limit ($N \to \infty$) depends on the decay rate of the interaction \cite{campa2003canonical}.  For a model in the strong long-range regime, $\alpha < d$, only a small number of modes have a non zero eigenvalue, such that $\lim_{N \to \infty} M/N = 0$. This can be seen analytically in models with a translation-invariant interaction matrix, which can be diagonalized in Fourier space, obtaining a closed expression for its eigenvalues: 
\begin{equation}
	D(\mathbf q) = \frac{\Gamma}{\tilde N} \sum_{\mathbf r} \tilde J(\mathbf r) \exp [-i \mathbf q \mathbf r] \,.
	\label{eq:eigenvalues}
\end{equation}
Here $\mathbf q$ denotes any of the $N$ reciprocal-space vectors in the first Brillouin zone and the sum runs over all lattice points. The large-$N$ behaviour of $D(\mathbf q)$ can then be estimated by replacing the sum with an integral \cite{campa2003canonical}. 
%
\begin{figure}
\centering
\includegraphics[width=\columnwidth] {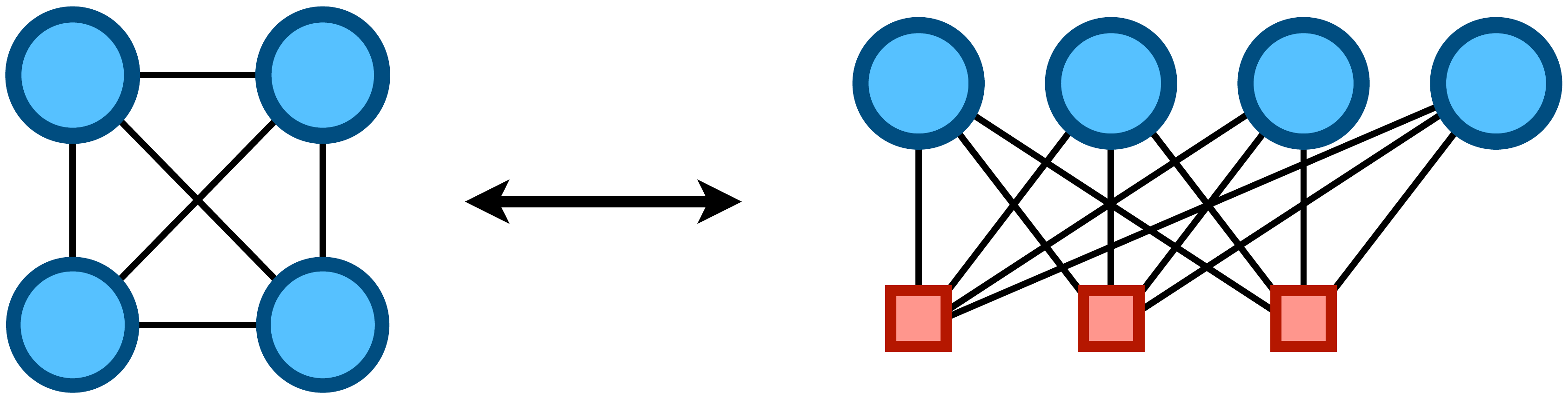}
\caption{Schematic of the generalized Hubbard-Stratonovich transformation mapping a quantum long-range model to (and from) a generalized Dicke model.  Blue dots represent ``matter'' degrees of freedom and red squares represent cavity modes.  Cf. Hamiltonians \eqref{eq:dickemodel} and \eqref{Heff} in the main text. }
\label{fig:generalHS}
\end{figure}

Complementarily, we provide in Fig. \ref{fig:levelstatistics} a graphical analysis of this phenomenon by showing the typical distribution of eigenvalues depending on $\alpha$ for $d=1$ (  the same behaviour is observed in other dimensions, not shown). This graphical analysis can be  useful for models without translation invariance. In Fig. \ref{fig:levelstatistics} we show that for a strong long-range model the eigenvalues bunch around zero as $N$ increases, whereas they remain more uniformly distributed in the weak long-range regime. This can be condensed into a criterion for determining whether arbitrary models are tractable: knowing that the eigenvalues of $J$ are non-negative and bounded by construction, if only a vanishingly small fraction $M/N$ are non-zero for $N\to\infty$, then their average will tend to zero and vice versa. Thus, for an arbitrary interaction matrix $J$, if 
\begin{equation}
    \lim_{N \to \infty} \frac{1}{N} \sum_{k=0}^{M-1} D_k = \lim_{N \to \infty} \frac{1}{N} {\rm Tr}(J) = 0
\end{equation}
the model is tractable, i.e. the number $M$ of non-zero eigenvalues scales as $\lim_{N \to \infty} M/N = 0$. If we apply this criterion to translation invariant models we find $\lim_{N \to \infty} 1/N \sum_k D_k = \lim_{N \to \infty} \Gamma b / \tilde N$, which is zero for $\alpha < 1$ and non-zero otherwise (See Apps. \ref{extensivity} and \ref{diagonalterms}.). 

Once it is established that a given model has a sufficiently small number of non-zero eigenvalues, one can sort them by decreasing value and truncate the sum  in Eq. \eqref{eq:Jij-decompositon-original} to consider only the first $M$ terms for which $D_k \neq 0$. For these remaining non-zero eigenvalues, we can identify $\omega_k = 1 / D_k$. This, together with the rescaled elements of the change of basis matrix,  $\lambda_{ik} = \sqrt{N} \Lambda_{ik}$, leads to Eq. \eqref{Heff} and effectively defines the mode structure for the associated Dicke model \eqref{eq:dickemodel}.

In summary, models for which $\lim_{N \to \infty} M/N = 0$, in particular strong long-range models, can be mapped to generalized Dicke models via the effective theory described in \cite{romanroche2021photon, romanroche2022effective}.  To gain further insight, it is useful to compare Hamiltonians \eqref{eq:dickemodel} and \eqref{Heff} with the system depicted in Figure \ref{fig:generalHS}, which outlines our procedure. The interacting model, shown in the left-hand side of the figure, where constituents are depicted as blue nodes and interactions as black edges, is mapped to a larger system where the \emph{physical} (matter) degrees of freedom are uncoupled and interact with \emph{auxiliary} bosonic modes represented as red squares \footnote{Fig. \ref{fig:generalHS} is actually an oversimplification, as it only depicts long-range interactions, which are the ones replaced by the auxiliary bosonic modes of the effective theory. The theory is also applicable to models containing a combination of short- and long-range interactions. For a discussion about the applicability of the method see Sec. \ref{sec:solvedicke}}. Integrating out these bosonic modes would lead back to the desired interactions $J_{ij}$. The parallelism between the auxiliary bosonic modes in the effective theory and the auxiliary classical fields in the standard HST motivates \emph{the claim of a generalized Hubbard-Stratonovich transformation}.


\section{Exact solution of strong long-range models}
\label{sec:solvedicke}

At this point we have shown how to map a strong long-range quantum model \eqref{eq:longrangecoupling} to a generalized Dicke model \eqref{eq:dickemodel} as illustrated in Fig. \ref{fig:generalHS}. To solve the latter, we will follow the steps outlined in the original solution of the Dicke model by Wang and Hioe \cite{wang1973phase, hioe1973phase}. In the thermodynamic limit, the trace over the photonic degrees of freedom is replaced by a collection of complex Gaussian integrals and the bosonic creation and annihilation operators, $a_k^\dagger$, $a_k$, are replaced by complex fields, $\alpha_k^*$, $\alpha_k$
\begin{widetext}
\begin{equation}
	Z = \int \prod_{k=0}^{M-1} \frac{d^2 \alpha_k}{\pi}  {\rm Tr}_{\rm m} \left\{ \exp \left[-\beta \left( \sum_{k=0}^{M-1} \omega_k |\alpha_k|^2 + \mathcal H_0 + \sum_{k, i} \frac{2 \lambda_{i k} x_k}{\sqrt{N}} \mathcal C_i \right) \right] \right\} \,,
	\label{eq:Zwang}
\end{equation}
\end{widetext}
where $\alpha_k = x_k + i y_k$. At this point, the parallelism with the standard Hubbard-Stratonovich transformation used in the classical model is even more explicit. The Gaussian integral over the imaginary parts $\{y_k\}$ yields an unimportant constant. To tackle the integration over the real parts, we perform a change of variables $u_k^2 = x_k^2 / N$ and define
\begin{equation}
\begin{split}
    Z_{\rm m}[u_k] &\equiv Z_{\rm m} (u_0, \ldots, u_{M-1}) \\
    &=  {\rm Tr}_{\rm m} \left\{ e^{-\beta \left( \mathcal H_0 + \sum_{k, i} 2 \lambda_{i k} u_k \mathcal C_i \right) } \right\}
\end{split}
\label{eq:Zmatter}
\end{equation}
and $f_{\rm m}[u_k] = \ln (Z_{\rm m}[u_k]) / N$. In the resulting integral
\begin{equation}
    Z = \int \prod_{k=0}^{M-1} \sqrt{\frac{N}{\pi \omega_k}} d u_k \exp \left(N \phi[u_k]\right) \,,
\end{equation}
where
\begin{equation}
	\phi[u_k] = -\beta \sum_{k=0}^{M-1} \omega_k u_k^2 + f_{\rm m}[u_k] \,,
\label{eq:exponentgeneral}
\end{equation}
the exponent depends explicitly linearly on $N$, allowing one to use the saddle-point method (exactly for $N \to \infty$) to express the partition function as the value of the integrand at the maximum of $\phi [u_k]$
\begin{equation}
    Z = \prod_{k=0}^{M-1} \sqrt{\frac{N}{\pi \omega_k}} \exp \left(N \phi[\bar u_k]\right) \,,
\label{eq:exactpartitionfunction}
\end{equation}
with,
\begin{equation}
\phi[\bar u_k] = \max_{\{u_k\}} \phi[u_k]  \,. 
\end{equation}
Computing the partition function is thus reduced to a multivariate maximization problem. In order for the zero-order saddle-point approximation to be exact, one has to verify that there exists a maximum $\{\bar u_k\}$, i.e. that $\phi$ admits a stationary point $\{\bar u_k\}$ and the eigenvalues of the Hessian of $\phi$ at the stationary point, $H_\phi[\bar u_k]$, are all negative. In the presence of several maxima, one has to find the global maximum. Finding global extrema of a multivariate scalar function is normally a complex task, without guarantee or provability of success, but in the present case it is greatly facilitated for homogeneous or near-homogeneous systems (See Sec. \ref{sec:ising}). Additionally, the second-order corrections to the partition function in the form of a factor $(\det H_\phi[\bar u_k])^{-1/2}$ must be negligible with respect to the zero-order term, $\exp \left\{N \phi[\bar u_k] \right\}$, but this is generally true, see App. \ref{secondordercorrections}. 
\emph{This is the main result of this paper, i.e. the exact expression for the partition function of strong long-range models \eqref{eq:exactpartitionfunction}}.

 In the next section and in order to give concrete formulas, we particularize for the case of the Ising model in transverse field.  
However, the ideas presented here can be applied to other models. For instance, our next section generalizes easily to a spin-$s$ system where $s>1/2$ and also to the inclusion of a longitudinal field, such that ${\mathcal H}_0 = \omega_z \sum_i S^z_i + \omega_x \sum_i S^x_i$ and $\mathcal C_i = 2 S_i^x$ with $[S_i^\alpha, S_j^\beta] = i \epsilon_{\alpha \beta \gamma} S^\gamma \delta_{ij}$ spin-$s$ operators. The Fermi-Hubbard model with long-range interactions could also be treated with our method. Here ${\mathcal H}_0 = t_{ij} c_i^\dagger c_j$ and ${\mathcal C}_i = c_i^\dagger c_i$ with $\{c_i, c_j^\dagger\} = \delta_{ij}$ fermionic operators. Finally, we could consider for $\mathcal H_0$ any model such that $\mathcal H_0 + \sum_i \xi_i \mathcal C_i$ is solvable, where the $\xi_i$ are constants. In doing so, we could combine short-range models (as the one-dimensional short-range Ising model in transverse field, the XY model, and so on) with strong long-range interactions. This is because our method requires knowledge of the eigenstates of $\mathcal H_0 + \sum_{k, i} 2 \lambda_{i k} u_k \mathcal C_i$, Cf. Eq. \eqref{eq:Zmatter}.

We have, thus far in the paper, focused only on ferromagnetic (attractive) models. However, the discussion of the applicability and generalizations of our method demands that we consider antiferromagnetic (repulsive) models  \cite{simon2011quantum, kaicher2023study}. Frustrated antiferromagnetic long-range models cannot be tackled with our method. To see why, it suffices to look at Fig. \ref{fig:levelstatistics}, frustrated antiferromagnetic models arise from changing the global sign of the interaction in  Eq. \eqref{eq:longrangecoupling}, which in turn results in a change of sign of the eigenvalues of the coupling matrix. For a general model, a shift to render the smallest eigenvalue equal to zero is not possible as it would require a $b$ of the order of $\tilde N$, leading to non-vanishing diagonal elements even in the $N \to \infty$ limit. For a model in which $\mathcal C_i^2 = 1$, the shift is possible, but after a shift to render the smallest eigenvalue equal to zero, we find that the majority of the eigenvalues are non-zero, regardless of the range of interactions $\alpha$. 
In contrast, it is possible to define unfrustrated long-range antiferromagnetic models, $J_{ij} = \Gamma (-1)^{i+j} \tilde J(\boldsymbol r_{ij}) / \tilde N$, as an extension of unfrustrated nearest-neighbour antiferromagnetic interactions \cite{gong2016topological}. Here, the sign change is alternating, rather than global, effectively defining two sublattices. The interaction matrix defined this way shares the eigenvalues of its ferromagnetic counterpart and the corresponding model can thus be tackled with our method. However, a number of interesting subtleties arise later on in the solution that deserve a detailed discussion, we reserve this for a future publication.


\section{Solution of the long-range Ising model in a transverse field.}
\label{sec:ising}
To showcase the effectiveness of the formalism presented in the previous sections, we particularize now to an Ising chain in transverse field 
\begin{equation}
    \mathcal H = \frac{\omega_z}{2} \sum_i^N \sigma^z_i - \sum_{ij}^N J_{ij} \sigma_i^x \sigma_j^x \,,
\end{equation}
where $\sigma^{x, z}$ are the usual Pauli matrices and $J_{ij}$ is given by Eq. \eqref{eq:Jalpha}. This corresponds to setting $d=1$, $\mathcal H_0 = \frac{\omega_z}{2} \sum_i \sigma^z_i$ and ${\mathcal C}_i = \sigma^x_i$. In App. \ref{wangsol} we show that this leads to 
\begin{equation}
	\phi[u_k] = -\beta \sum_{k=0}^{M-1} \omega_k u_k^2 + \frac{1}{N} \sum_i^N  \ln\left[2 \cosh (\beta \epsilon_i) \right]  \,,
\label{eq:exponent}
\end{equation}
with $2 \epsilon_i [u_k] = \sqrt{\omega_z^2 + 4 \left(2 \sum_k \lambda_{i k} u_k \right)^2}$.

In the case of a homogeneous ${\mathcal H}_0 $  we show in Apps. \ref{homogeneousmax} and \ref{globalmax} that the global maximum is homogeneous in the lattice. In terms of the minimization variables, homogeneity implies that $u_0 = u \neq 0$ and $u_{k \neq 0} = 0$, see  App. \ref{homogeneousmax}. This means that only the zero mode, which is constant on the lattice, $\lambda_{i0} = 1 \; \forall i$, is relevant in determining the thermodynamic properties of the model. In turn, one finds that the critical properties of the model are independent of the decay rate of interactions $\alpha$, since the latter only determines the degree to which higher-frequency modes ($k > 0$) have to be considered in the diagonalization of $J$.  In more intuitive terms, homogeneity is revealed in the fact that $2 \epsilon_i = 2 \epsilon = \sqrt{\omega_z^2 + 16 u^2}, \; \forall i$. In any case, the multivariate maximization problem simplifies to a single variable maximization problem $\max_{u} \phi(u)$. Taking the derivative of $\phi$ with respect to $u$ yields the condition
\begin{equation}
    \bar u \bar \epsilon = 2 \Gamma \bar u \tanh \left(\beta \bar \epsilon \right) \,,
    \label{eq:homogeneoussolution}
\end{equation}
which is manifestly $\alpha$-independent. Note that $\bar \epsilon = \epsilon(\bar u)$. For $\Gamma < \omega_z/4$, $\bar u = 0$ is the only solution. For $\Gamma > \omega_z/4$, the solution depends on $\beta$, for $\beta > \beta_c$, with $\beta_c$ given by $\omega_z = 4 \Gamma \tanh \left(\beta_c \omega_z / 2 \right)$, there is another solution to Eq. \eqref{eq:homogeneoussolution} given by $\bar \epsilon = 2 \Gamma \tanh \left(\beta \bar \epsilon \right)$. The solution $\bar u = 0$ corresponds to a maximum in the regime where it is the only solution and becomes a minimum for $\beta > \beta_c$ with the maximum given by the other solution \cite{wang1973phase}.  This marks the paramagnetic-ferromagnetic transition point. This is the well-known mean-field critical behaviour of the standard (single-mode homogeneous coupling) Dicke model \cite{lieb1973the, wang1973phase}, which is shared by the Lipkin-Meshkov-Glick (LMG) model (all-to-all homogeneous Ising) \cite{brankov1975asymptotically, gibberd1974equivalence} and, as we just showed, is also universal to all strong-long-range Ising models and their associated Dicke models,  \emph{i.e.} we have demonstrated that the critical point is independent of $\alpha$.  This can be visualized in Fig. \ref{fig:observables} where the vertical line marks the phase transition, located at the maximum for the susceptibility (see below), and is independent of $\alpha$.
Besides, in Fig. \ref{fig:phasediagram} we compute the critical line, in red, in the $(\Gamma, 1/\beta)$-plane and compare it against the simulations in  Ref. \cite{gonzalezlazo2021finitetemperature}. We find excellent agreement with their numerical results and showcase that the critical point is independent of $\alpha$ and coincides with the mean-field value.

In terms of observables, we focus now on the calculation of magnetization.
In order to do so from the partition function, we introduce a perturbative longitudinal field to the Hamiltonian, such that $\mathcal H \to \mathcal H - \sum_i h_i \sigma^x_i$. Then one can compute the order parameter $\beta \langle\sigma^x_i\rangle = \partial \ln Z/ \partial h_i$ and the susceptibilities 
\footnote{The magnetization must be kept $\{h_n\}$-dependent in order to compute the susceptibility, the magnetization of the $\{h_n\}$-independent model is defined as $\beta \langle\sigma^x_i\rangle = \lim_{\{h_n\} \to 0} \partial \ln Z/ \partial h_i$}
\begin{equation}
\chi_{ij} = \lim_{\{h_n\} \to 0} 
\frac{\partial \langle\sigma^x_i\rangle}{\partial h_j}
\, .
\label{eq:chiij}  
\end{equation}
The introduction of longitudinal fields leads to the substitution $2 \sum_k \lambda_{i k} u_k \to 2 \sum_k \lambda_{i k} u_k + h_i$ in Eq. \eqref{eq:exponent}. The magnetization is then
\begin{equation}
	\langle\sigma^x_i\rangle = \tanh \left(\beta \epsilon_i[\bar u_k]\right) \frac{2 \sum_{k=0}^{M-1} \lambda_{i k} \bar u_k + h_i}{\epsilon_i [\bar u_k]} \,.
	\label{eq:magnetization}
\end{equation}
Here, the magnetization appears as a function of the maximization variables $\{\bar u_k\}$. However, it is possible to show that $\sum_k \lambda_{ik} \bar u_k = \sum_{j} J_{ij} \langle \sigma_j ^x \rangle$, rewriting Eq. \eqref{eq:magnetization} as a self-consistent equation on $\{\langle \sigma_i^x \rangle\}$ which is precisely the self-consistent equation that arises in a mean-field solution. Our exact analytical method is thus equivalent to mean-field theory, proving that mean-field theory is exact for strong long-range models and any lattice dimensionality $d$. Anecdotally, our theory evidences that the self-consistent solution from mean-field theory is redundant, in the sense that the solution involves a transcendental equation of $N$ variables (the magnetizations $\{\langle \sigma_i^x \rangle\}$), whereas the same problem can be rewritten in terms of $M$ variables (the $\{u_k\}$), with $\lim_{N \to \infty} M/N = 0$.


\subsection{Decay of correlations}
\label{correlationdecay}
\begin{figure}
    \centering
    \includegraphics[width = 1.0 \columnwidth]{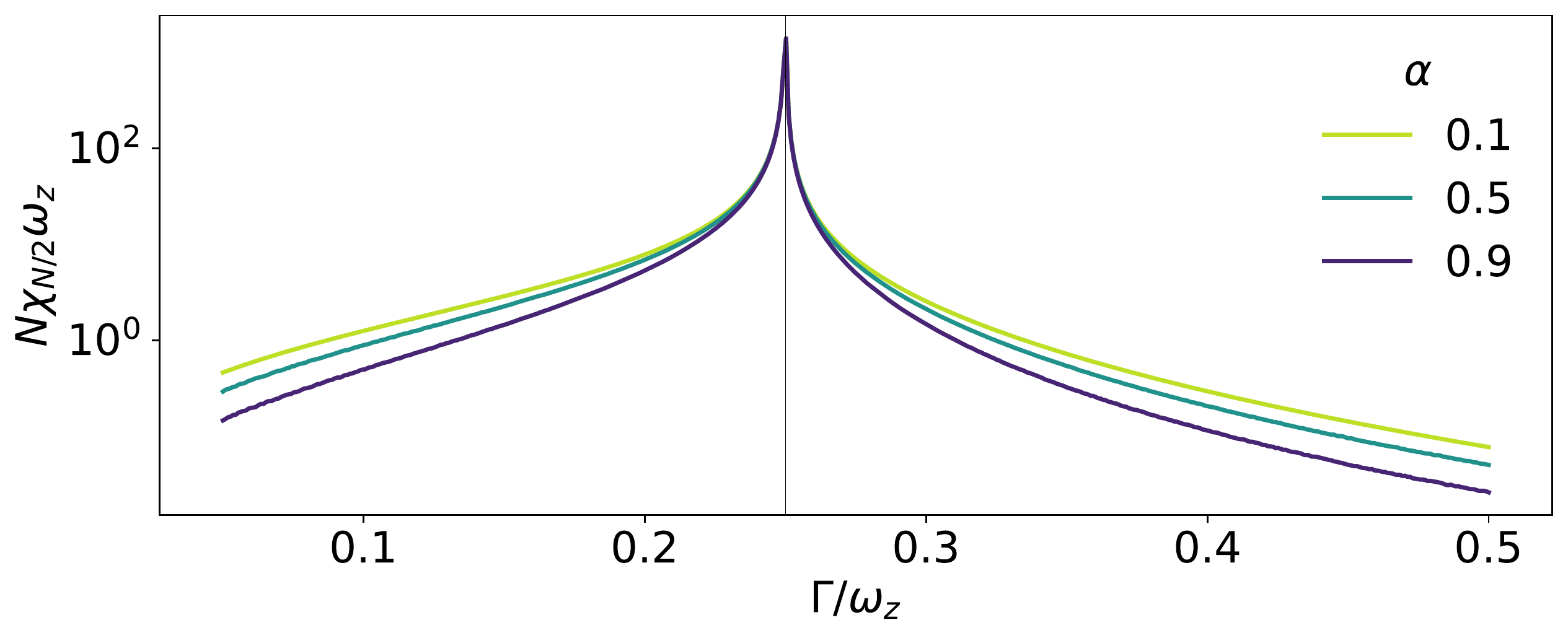}
    \caption{Susceptibility for maximally separated spins $\chi_{N/2}$ for strong long-range quantum Ising models as a function of the interaction strength $\Gamma$. The parameters used were $\omega_z = 1$, $\beta \omega_z = 10$ and $N = 100$.}
    \label{fig:observables}
\end{figure}
Our final result concerns the decay of correlations. In weak long-range and short-range systems correlations decay exponentially at long distances. Only at the critical point do these systems exhibit power law decay of correlations \cite{vodola2015longrange, vanderstraeten2018quasiparticles, francica2022correlations}. Conversely, strong long-range systems exhibit power law decay of correlations at all distances. In the absence of exponential decay, the concept of correlation length cannot be straightforwardly defined, although there have been some attempts \cite{sadhukhan2021correlation}. Here we study the susceptibility $\chi_{ij}$ \eqref{eq:chiij} as a measure of correlations between spins, as it is proportional to the Kubo correlator \cite{kubo1991statistical}[Chap. 4]
\begin{equation}
    \chi_{ij} = \beta \left(\frac{1}{Z} \int_0^\beta {\rm Tr} \left(e^{-(\beta - s)\mathcal H} \sigma_i^x e^{-s \mathcal H} \sigma_j^x\right) - \langle\sigma^x_i\rangle\langle\sigma^x_j\rangle\right) \,.
\label{eq:kubo}
\end{equation}
The susceptibility can be computed analytically from Eq. \eqref{eq:chiij} for a translation invariant model \cite{schwabl}[Chap. 6] (See. App. \ref{susceptibility} for a derivation) or numerically otherwise. The analytical derivation yields
\begin{equation}
    \chi_{ij} = Y \delta_{ij} + \frac{1}{N} \sum_{k=0}^{M-1} \lambda_{ij} \left( \chi_k - Y \right) \lambda_{jk} \,,
    \label{eq:chianal}
\end{equation}
where $Y$ is a quantity that depends on $\bar u$ \eqref{eq:Y} and $\{\chi_k\}$ are the Fourier modes of the susceptibility \eqref{eq:chieqfourierspace}. Eq. \eqref{eq:chianal} evidences that $\chi_{i \neq j}$ goes to zero in the thermodynamic limit with a speed that is determined by the ratio $M/N$ and thus ultimately by $\alpha$ (by its relation to $d$).

For a numerical calculation, the introduction of a site-dependent field $h_i$ breaks the homogeneity of the model and the multivariate maximization of $\phi$ is carried out numerically, $\langle\sigma^x_i\rangle$ is then computed according to Eq. \eqref{eq:magnetization} and $\chi_{ij}$ is computed as a finite difference. We have verified that both methods yield the same results for the current model. This is noteworthy because the numerical calculation relies on a multivariate optimization which could, a priori, converge to an incorrect result corresponding to a local maxima. We believe the success is due to the fact that the only deviation from homogeneity stems from the introduction of a perturbative field and is thus small. Hence, although the optimization is strictly multivariate, the landscape does not differ much from the univariate case.

Despite the fact that the analytical results have been obtained under the assumption that we worked in the thermodynamic limit $N \to \infty$. The computation of the susceptibility, whether numerically or according to Eq. \eqref{eq:chianal}, requires us to fix a finite value of $N$ and $M$. For each value of $\alpha$ and $N$, we increase the value of $M$ until convergence is reached while enforcing the constraint that $\lim_{N \to \infty} M(N)/N = 0$.

\begin{figure}
    \centering
    \includegraphics[width=\columnwidth]{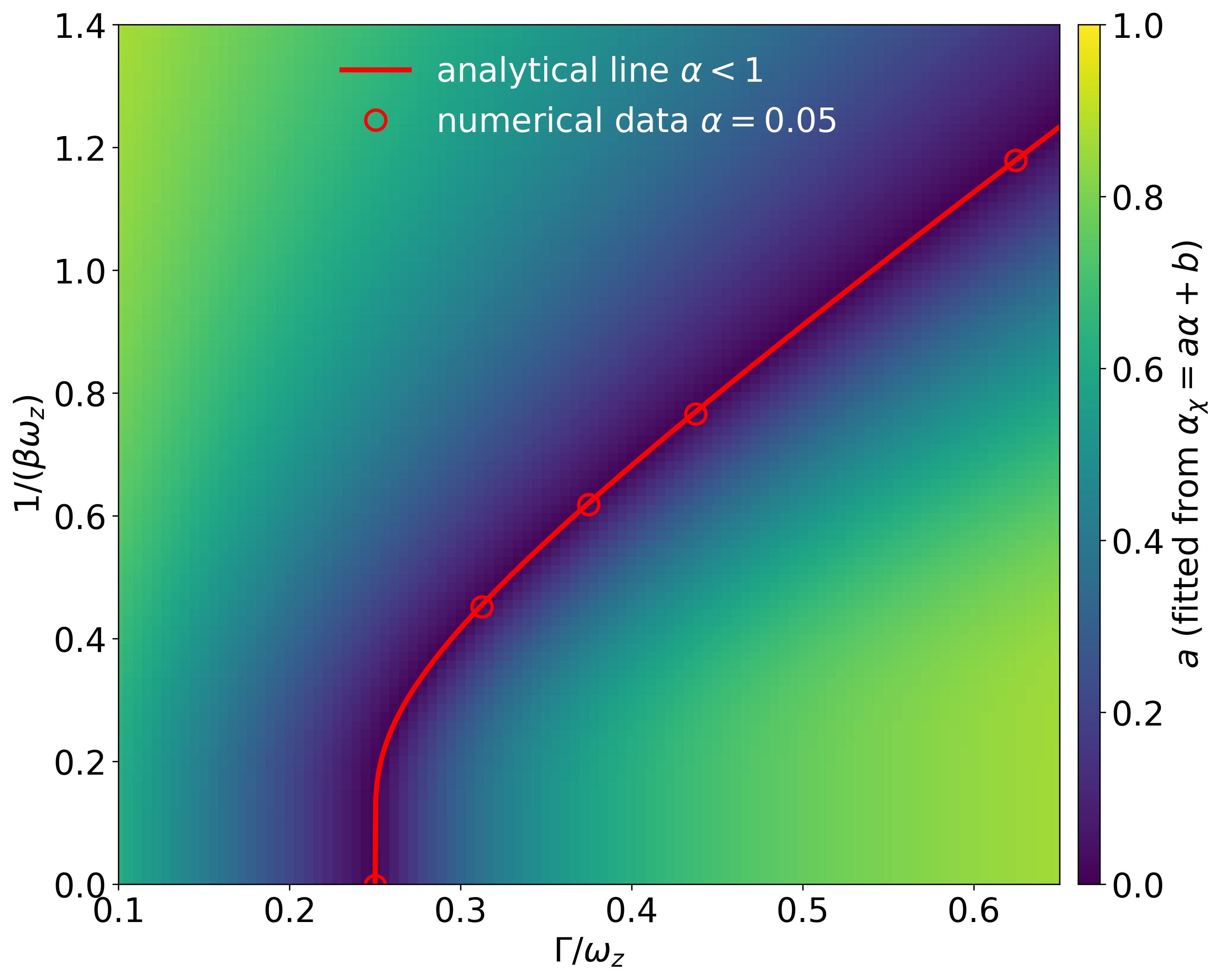}
    \caption{Phase diagram of the strong long-range Ising chain in transverse field. The red line corresponds to the universal critical line for $\alpha < 1$. The numerical data is taken from Ref. \cite{gonzalezlazo2021finitetemperature}. The colormap shows the slope of the linear dependence $\alpha_\chi = a \alpha + b$, where $\alpha_\chi$ is susceptibility decay rate and $\alpha$ is the rate of decay of interactions, computed with parameters $\omega_z = 1$ and $N = 100$.}
    \label{fig:phasediagram}
\end{figure}

Because the model is translation invariant, the susceptibility is only a function of distance, allowing us to define $\chi_{i i+r} \equiv \chi_r$. In Fig. \ref{fig:observables} we study the susceptibility at a fixed distance: we plot the half-chain susceptibility $\chi_{N/2}$ as a function of the interaction strength $\Gamma$ for different decay rates $\alpha < 1$ at zero temperature $\beta \to \infty$. The half-chain susceptibility displays $\alpha$-independent divergence at the critical point and some dependence on $\alpha$ away from it. Intuitively, the correlations remain larger for longer-ranged models. We now turn to the spatial dependence of the susceptibility. In the absence of a correlation length, we study the susceptibility decay rate, $\alpha_\chi$, defined from the relation $\chi_r = A \cdot r^{-\alpha_\chi} $. Interestingly, one finds that $\alpha_\chi$ depends linearly on $\alpha$, $\alpha_\chi = a \alpha + b$. In Fig. \ref{fig:phasediagram} we plot the slope $a$ as a function of interaction strength $\Gamma$ and inverse temperature $1/\beta$. Close to the critical line, the susceptibility decay rate $\alpha_\chi$ becomes independent of the interaction decay rate $\alpha$, in agreement with Fig. \ref{fig:observables}. As one moves further from the critical line, $a \to 1$, varying continuously from $0$ to $1$ in intermediate regions. In all cases we find $b \approx 0$. Similar algebraic decays have been described previously for the connected correlator $\langle \sigma_1^x \sigma_r^x \rangle - \langle \sigma_1^x \rangle \langle \sigma_r^x \rangle$ in the paramagnetic phase \cite{vodola2015longrange}. There, a linear relation between $\alpha_\chi$ and $\alpha$ is also reported. Here we extend those findings to the full phase diagram. 

\section{Conclusions}
\label{conclusions}

In this paper, we have presented a method for solving strong long-range models in the quantum domain based on the Hubbard-Stratonovich transformation for classical systems. Our method is a physics-based solution rooted in  light-matter interaction models in which, in the thermodynamic limit, light can be integrated out leaving an effective long-range model.   Solutions of the former, \emph{i.e.} Dicke models, are due to  Hepp and Lieb \cite{Hepp1973, Hepp1973a}, and Wang and Hioe \cite{wang1973phase, hioe1973phase}, which we have recently generalized \cite{romanroche2021photon, romanroche2022effective}.

We have shown that our method can be applied in the strong long-range regime and
confirmed that mean field theory is exact in this regime. Cf. Fig. \ref{fig:longcla}  Eqs. \eqref{eq:Jalpha}, \eqref{eq:dickemodel}  and \eqref{Heff}.  In doing so,  this paper  complements the work of Mori \cite{mori2012equilibrium}. Besides, it  extends the work of Campa and coworkers for classical strong long-range models to the quantum case \cite{campa2003canonical}. It is worth noting that neither our method nor the equivalent mean field theory can be used to compute non-local quantities such as the entanglement entropy, which can be non-trivial in strong long-range systems \cite{latorre2005entanglement}.

Our method is flexible and could be applied e.g. to spin-$s$ systems where $s>1/2$, to models with a longitudinal field, such as the long-range XXZ model, or to the Fermi-Hubbard model with long-range interactions. Additionally, many exactly solvable models could be complemented with long-range interactions and solved with our method, since it relies on knowing the eigenvalues of the system without long-range interactions and with a ``field'' term proportional to the long-range coupling operator. Unfrustrated antiferromagnetic systems are also within the scope of the method.
In conclusion, our work provides a new and powerful tool for solving quantum long-range models.

\section*{Acknowledgements}
We thank Alessandro Campa and Takashi Mori for discussing and explaining details of their previous results, they were of great help in completing this work.
The authors acknowledge  funding from the EU (QUANTERA SUMO and FET-OPEN Grant 862893 FATMOLS), the Spanish Government Grants PID2020-115221GB-C41/AEI/10.13039/501100011033 and TED2021-131447B-C21 funded by MCIN/AEI/10.13039/501100011033 and the EU ``NextGenerationEU''/PRTR, the Gobierno de Arag\'on (Grant E09-17R Q-MAD) and the CSIC Quantum Technologies Platform PTI-001.
This work has been financially supported by the Ministry of Economic Affairs and Digital Transformation of the Spanish Government through the QUANTUM ENIA project call - Quantum Spain project, and by the European Union through the Recovery, Transformation and Resilience Plan - NextGenerationEU within the framework of the ``Digital Spain 2026 Agenda''. J R-R acknowledges support from the Ministry of Universities of the Spanish Government through the grant FPU2020-07231.
\appendix

\section{Properties of the long-range interaction matrix}
\subsection{Loss of extensivity}
\label{lossextensivity}
\label{extensivity}
In Fig. \ref{fig:b} we illustrate the extensivity (or lack thereof) of a model with power-law decaying interactions in $d=1$. We compute $\tilde N = \sum_j \tilde J_{ij}$ as a measure of the coupling energy per spin. In the absence of Kac's rescaling, this quantity must not scale with the number of spins $N$ to keep the total coupling energy extensive. Fig. \ref{fig:b} shows that this is not the case for $\alpha <1$. The threshold case $\alpha =1$ is highlighted for clarity and corresponds to a logarithmic dependence of $\tilde N$ on $N$. For $\alpha > 1$ the dependence is sublogarithmic, i.e. $\tilde N$ becomes independent of $N$ at large $N$. As discussed in the main text, loss of extensivity is prevented with Kac's rescaling, which we can now understand as a renormalization of the total coupling energy by the energy per spin.

In fact, for $d=1$ and $N \to \infty$ the scaling of $\tilde N$ can be computed analytically, since
\begin{equation}
    \tilde N = \sum_i^N \tilde J_{ij} = b + 2\sum_{r=1}^\infty r^{-\alpha} \; .
\end{equation}
As $b$ converges to a constant value when $N\to\infty$ (will be shown in App. \ref{diagonalterms}), the convergence of $\tilde N$ will be ruled by the convergence of the infinite series. For $\alpha > 1$ the series is convergent, so $\tilde N$ becomes independent of $N$ at large $N$. For $\alpha \leq 1$ the series diverges.

\subsection{Diagonal terms can be neglected in strong-long range models}
\label{diagonalterms}
Setting $b \neq 0$ introduces a new diagonal term in the Hamiltonian
\begin{equation}
	-\sum_i^N \Gamma b/\tilde N \mathcal C_i^2 \,.
\end{equation}
Importantly, this term contains a factor $b/\tilde N$. We know from App. \ref{extensivity} that $\lim_{N \to \infty} \tilde N = \infty$ for $\alpha < d$. So if $b$ is independent of $N$ for $N \to \infty$, the diagonal term vanishes in the thermodynamic limit. 

It can be shown analytically that this is the case for $d=1$ and $\alpha > 0$. From Eq. \eqref{eq:eigenvalues} we see that the smallest eigenvalue of $J$ when $N \to \infty$ is given by
\begin{equation}
    D_{\rm min} = \frac{\Gamma}{\tilde N} \sum_{\mathbf{r}} \tilde J \left( \mathbf{r} \right) \left( -1 \right)^{r} = \frac{\Gamma}{\tilde N} \left( b + 2\sum_{r=1}^{\infty} \left(-1\right)^r r^{-\alpha} \right) \,.
\end{equation}
Here $r = |\mathbf r|$. Hence, setting $D_{\rm min} = 0$ fixes 
\begin{equation}
    b = -2 \sum_{r=1}^\infty (-1)^r r^{-\alpha} \,.
\end{equation}
The convergence of this series is proven by means of the alternating series test, since the absolute value of its terms monotonically decrease to 0. For $\alpha = 0$ and finite $N$, as $\tilde J_{i\neq j} = 1$, $b$ must be fixed to $1$ to ensure that the smallest eigenvalue of $\tilde J$ is zero. This is manifestly independent of $N$.

\begin{figure}
    \centering
    \includegraphics[width = \columnwidth]{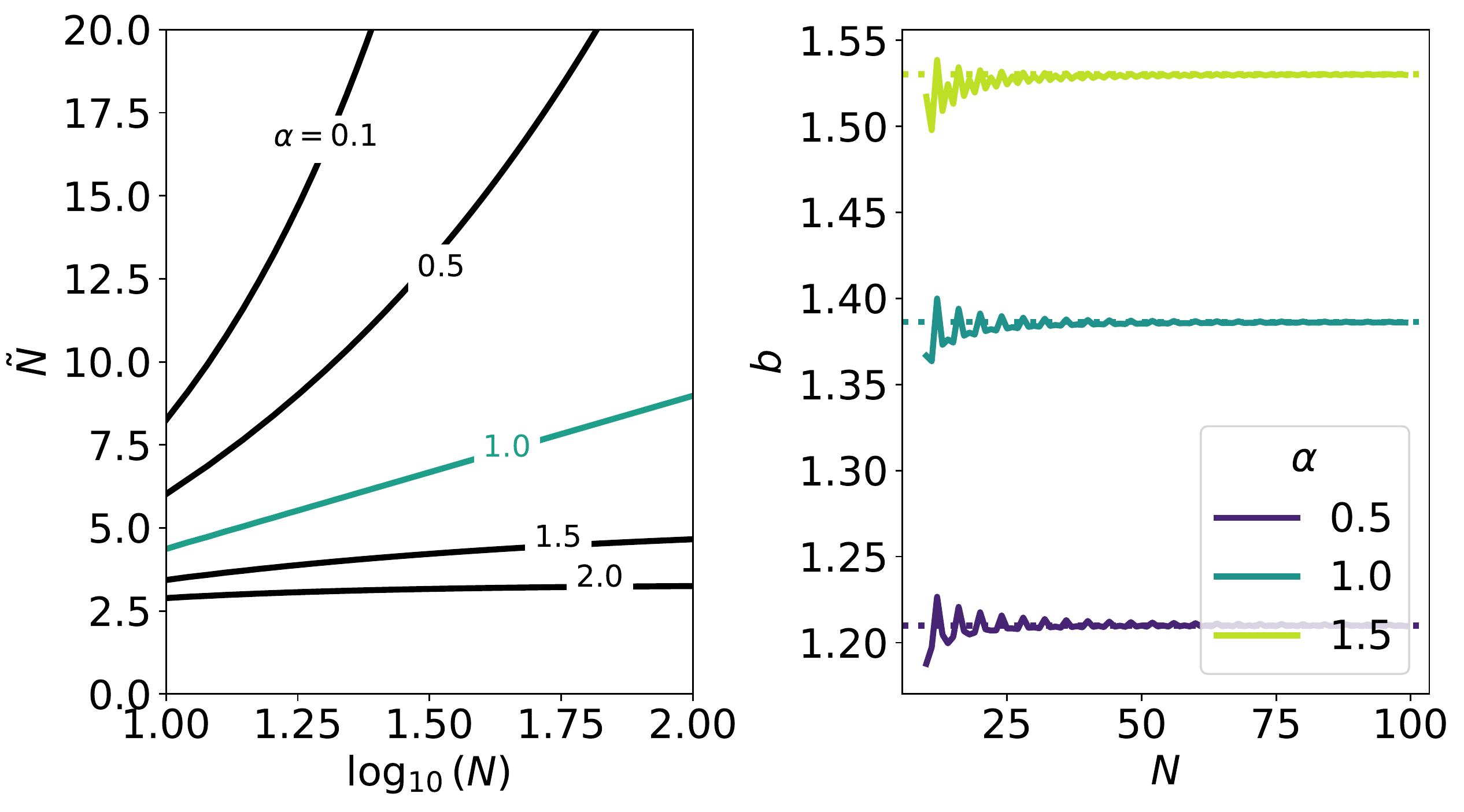}
    \caption{Scaling of $\tilde N = \sum_j^N \tilde J_{ij}$ (left) and $b$ (right) for a one dimensional model ($d=1$) with power-law decaying interactions \eqref{eq:longrangecoupling}. Here $b$ is fixed such that the smallest eigenvalue of $\tilde J$ is zero. The dotted lines indicate the analytical asymptote.}
    \label{fig:b}
\end{figure}

In other dimensions or other models, this test can be done graphically. In Fig. \ref{fig:b} we show the value of $b$, for a model with power-law decaying interactions in $d=1$, computed numerically for different values of $N$ when $b$ is chosen so that the smallest eigenvalue of $J$ is zero. One can see that in this case the numerical results converge to the analytical prediction. The same behaviour is observed in other dimensions.

\section{Negligible second order corrections to the saddle point method}
\label{secondordercorrections}
The second order term of the saddle-point expansion is proportional to $(\det H_\phi[\bar u_k])^{-1/2}$. Accordingly, it corresponds to a correction to the free energy per particle of the form 
\begin{equation}
	\frac{1}{N} \ln(\det H_\phi[\bar u_k]) = \frac{1}{N} \sum_{k=0}^{M-1} \ln \nu_k \propto \frac{M}{N} \,,
\end{equation}
where the $\{\nu_k\}$ are the eigenvalues of $\det H_\phi[\bar u_k]$. This correction scales as $M/N$ and thus vanishes in the thermodynamic limit. Notably, if the applicability of the effective theory to map a long-range interacting model to a generalized Dicke model constitutes the first appearance of the restriction $\lim_{N \to \infty} M/N = 0$, the argument contained in this Appendix constitutes a second independent one. In fact, this second occurrence of the restriction also appears in classical systems, where it is actually the only restriction to $\lim_{N \to \infty} M/N = 0$, as in classical systems an unrestricted standard HST can be used, as outlined in Sec. \ref{sec:sketch}.

\section{Solving the long-range Ising model in transverse field}
\subsection{Solving the associated Dicke model}
\label{wangsol}

Particularizing Eq. \eqref{eq:Zmatter} for the Ising model, we have to compute
\begin{equation}
\begin{split}
    Z_{\rm m}[u_k] = {\rm Tr}_{\rm m} \left\{ \exp \left[-\beta \sum_i^N \Biggl( \right. \right. & \frac{\omega_z}{2} \sigma_i^z \\
    &\left. \left. \left. + \sum_{k=0}^{M-1} 2 \lambda_{i k} u_k \sigma_i^x \right) \right] \right\} \,.
\end{split}
\end{equation}
Because the spins are now decoupled, the trace over spins factorizes. The resulting single spin Hamiltonian has eigenvalues 
\begin{equation}
	\epsilon_i^\pm = \pm \epsilon_i = \pm \frac{1}{2} \sqrt{\omega_z^2 + 4  \left(\sum_{k=0}^{M-1} 2 \lambda_{i k} u_k \right)^2} \,.
\end{equation}
Accordingly,
\begin{align}
    & Z_{\rm m}[u_k] = \prod_i 2 \cosh (\beta \epsilon_i) \,, \\
    & f_{\rm m}[u_k] = \frac{1}{N} \sum_i \ln(2 \cosh \beta \epsilon_i) \,, \\
    & \phi_{\rm m}[u_k] = -\beta \sum_k \omega_k u_k^2 +\frac{1}{N} \sum_i \ln(2 \cosh \beta \epsilon_i) \,.
\end{align}

\subsection{Existence of a homogeneous maximum of $\phi$}
\label{homogeneousmax}
To find the maximum of $\phi[u_k]$ we impose a vanishing gradient: $\nabla \phi = 0$, which translates to the following condition for the maximization variables
\begin{equation}
    \bar u_k \omega_k = \frac{1}{N} \sum_i^N \tanh(\beta \bar \epsilon_i) \frac{\lambda_{ik} 2 \sum_{l=0}^{M-1} \lambda_{il} \bar u_l}{\bar \epsilon_i} \,.
    \label{eq:extremumcondition}
\end{equation}
From here, let us consider a solution that is homogeneous in the lattice, we will later prove that possible inhomogeneous maxima are only local maxima \ref{globalmax}. Let us define $\mu_i = 2 \sum_k \lambda_{ik} u_k$ and consider it as an alternate optimization variable. Homogeneity implies that $\bar \mu_i \equiv \bar \mu$, to see how this affects the variables $\{\bar u_k\}$ it is useful to invert the relation and write $u_k$ in terms of the $\{\mu_i\}$, yielding 
\begin{equation}
    u_k = \frac{1}{2N} \sum_i^N \lambda_{ik} \mu_i \,.
\end{equation}
Now, homogeneity implies
\begin{equation}
    \bar u_k = \frac{\bar \mu}{2N} \sum_i^N \lambda_{ik} \,.
\end{equation}
Since the $\{\lambda_{ik}\}$ are the Fourier modes resulting from the diagonalization of $J$, we have $\sum_i \lambda_{ik} = N \delta_{k0}$. Accordingly, we find $\bar u_{k \neq 0} = 0$ and $2 \bar u_0 = \bar \mu$.
So, if the solution is homogeneous, the only relevant mode is the zero mode $\bar u_0 \equiv \bar u$ and the rest of the maximization variables are zero, with $\bar u$ satisfying the condition
\begin{equation}
    \bar u \omega_0 = \frac{2\bar u}{\bar \epsilon} \tanh(\beta \bar\epsilon) \,.
    \label{eq:homogeneoussolutionapp}
\end{equation}
From Eq. \ref{eq:eigenvalues} we have $\omega_0 = 1/\Gamma$, which when replaced in Eq. \eqref{eq:homogeneoussolutionapp} yields Eq. \eqref{eq:homogeneoussolution}.

At this point we can compute the Hessian of $\phi$, $H_\phi$. As we have shown, for a homogeneous solution, the optimization problem becomes single-valued such that
\begin{equation}
\begin{split}
 	H_\phi =  \frac{\partial^2 \phi}{\partial u^2}  = & -2 \beta \omega_0 + \left[1 - \tanh^2(\beta \epsilon)\right] \left(\frac{4 \beta u}{\epsilon} \right)^2 \\
	&+ \beta \tanh(\beta \epsilon) \left( \frac{4}{\epsilon} - \frac{16u^2}{\epsilon^3} \right) \,.
\end{split}
\end{equation}
If we evaluate the Hessian at $\bar u=0$, which is always a solution of Eq. \ref{eq:homogeneoussolutionapp} we obtain 
\begin{equation}
	H_\phi(\bar u = 0) = -2 \beta \omega_0 + \beta \tanh\left(\beta\frac{\omega_z}{2}\right) \frac{8}{\omega_z} \,,
\end{equation}
which is always negative for $\Gamma < \omega_z/4 $, i.e. for $\omega_0 = 1 / \Gamma > 4 / \omega_z$. For $\omega_0 > 4 / \omega_z$, the sign depends on $\beta$, being negative for $\beta < \beta_c$ with $\omega_0 \omega_z = 4 \tanh \left(\beta_c \omega_z / 2 \right)$. So in the regime where $\bar u = 0$ is the only solution to Eq. \eqref{eq:homogeneoussolutionapp}, it is a maximum. For  $\omega_0 < 4/\omega_z$ and $\beta > \beta_c$ a non-trivial solution given by $\omega_0 \bar \epsilon = 2 \tanh(\beta \bar \epsilon)$ appears and can be shown graphically to be the maximum. Therefore, Eq. \eqref{eq:homogeneoussolutionapp} always has a solution that is a maximum of $\phi$. 

\subsection{Proof that the homogeneous solution is the global maximum}
\label{globalmax}

We cannot rule out the existence of inhomogenoeus extrema of $\phi$, i.e. inhomogeneous solutions of Eq. \eqref{eq:extremumcondition}. Instead, we show that if there exists an inhomogeneous solution and it is a maximum, it is a local maximum, with the global maximum given by the homogeneous solution.

Let us express the self-consistent condition for the extrema of $\phi$ given in Eq. \eqref{eq:extremumcondition} in terms of $\{\mu_i\}$ and $\{\epsilon_i\}$
\begin{equation}
	\frac{1}{2N} \sum_i^N \lambda_{ik} \bar \mu_i \omega_k = \frac{1}{N} \sum_i^N \tanh(\beta \bar \epsilon_i) \frac{\lambda_{ik} \bar \mu_i}{\epsilon_i} \,.
\end{equation}
Isolating the $\{\bar \mu_i\}$ yields the self-consistent condition
\begin{equation}
	\bar \mu_i = 2 \sum_j^N \tanh(\beta \bar \epsilon_j) \frac{\bar \mu_j}{\bar \epsilon_j} J_{ij} \,,
	\label{eq:selfconsistentmu}
\end{equation}
which is simply a reformulation of the maximization problem in terms of new variables. Accordingly, $\phi$ reads
\begin{equation}
	\phi[\mu_i] = - \frac{\beta}{4N} \sum_{ij}^N  \mu_i J^+_{ij} \mu_j + \frac{1}{N} \sum_i^N \ln\left[2 \cosh (\beta \epsilon_i) \right] \,,
	\label{eq:phimu}
\end{equation}
with $N J^+_{ij} = \sum_k \lambda_{ik} \omega_k \lambda_{jk}$. Note that $J J^+ J = J$.
Substituting Eq. \eqref{eq:selfconsistentmu} in Eq. \eqref{eq:phimu} yields
\begin{equation}
\begin{split}
	\phi[\bar \mu_i] &= - \frac{\beta}{2N} \sum_{i}^N \frac{\tanh(\beta \bar \epsilon_i)}{\bar \epsilon_i} \bar \mu_i^2  + \frac{1}{N} \sum_i^N \ln\left[2 \cosh (\beta \bar \epsilon_i) \right] \\
	&= \frac{1}{N} \sum_i^N \xi(\bar \mu_i) \,.
\end{split}
\end{equation}
We can particularize this expression for the homogenous solution, $\bar \mu_i \equiv \bar \mu$,
\begin{equation}
\begin{split}
	\phi(\bar \mu) & = - \frac{\beta}{2} \frac{\tanh(\beta \bar \epsilon)}{\bar \epsilon} \bar \mu^2  +  \ln\left[2 \cosh (\beta \bar \epsilon) \right] \\
	& = \frac{1}{N} \sum_i^N \xi(\bar \mu) \,.
\end{split}
\end{equation}
Note that because $\bar \mu$ maximizes $\phi$, it also maximizes $\xi$. Therefore, an inhomogeneous maximum of $\phi$ given by $\{\bar \mu_i\}$ cannot maximize $\xi$ for all $\bar \mu_i$ (to the extent that some $\bar \mu_i$ must deviate from $\bar \mu$ in order for the configuration to be inhomogeneous) and thus $\phi(\bar \mu)\geq \phi[\bar \mu_i]$. The global maximum of $\phi$ is given by the homogeneous solution.

\section{Analytical calculation of susceptibilities}
\label{susceptibility}
From Eqs. \eqref{eq:magnetization} and \eqref{eq:extremumcondition} we realize that 
\begin{equation}
    \bar u_k \omega_k = \frac{1}{N} \sum_i^N \langle \sigma_i^x \rangle \lambda_{ik}
\end{equation}
and thus
\begin{equation}
    \frac{\partial \bar u_k}{\partial h_j} \omega_k = \frac{1}{N} \sum_i^N \chi_{ij} \lambda_{ik} \,.
    \label{eq:ukder}
\end{equation}
From Eq. \eqref{eq:chiij} we have
\begin{equation}
    \chi_{ij} = \lim_{\{h_n\} \to 0} \left(2 \sum_{k=0}^{M-1} \lambda_{ik} \frac{\partial \bar u_k}{\partial h_j} + \delta_{ij} \right) Y_i \,,
    \label{eq:chianaldef}
\end{equation}
with
\begin{equation}
\begin{split}
    Y_i = &\left(1 - \tanh^2(\beta \bar \epsilon_i) \right) \beta \left(\frac{\bar \mu_i + h_i}{\bar \epsilon_i} \right)^2 \\
    &+ \tanh(\beta \bar \epsilon_i) \frac{1}{\bar \epsilon_i} \left[1  - \left(\frac{\bar \mu_i + h_i}{\bar \epsilon_i} \right)^2 \right] \,.
    \label{eq:Y}
\end{split}
\end{equation}
From Eqs. \eqref{eq:ukder} and \eqref{eq:chianaldef} and after some manipulation, we find
\begin{equation}
    \chi_{ij} = \left(2 \sum_r^N J_{ir} \chi_{rj} + \delta_{ij}\right) Y \,,
    \label{eq:chieqrealspace}
\end{equation}
where $Y = \lim_{\{h_n\} \to 0} Y_i$. For a translation-invariant model, Eq. \eqref{eq:chieqrealspace} can be solved in Fourier space. We define
\begin{equation}
    \chi_{ij} = \frac{1}{N} \sum_{k=0}^{N-1} \lambda_{ik} \chi_k \lambda_{jk}
\end{equation}
and find
\begin{equation}
    \chi_k = \frac{Y}{1 - 2 Y D_k} \,.
    \label{eq:chieqfourierspace}
\end{equation}
Here $D_k$ are the eigenvalues of $J$, Cf. Eq. \eqref{eq:Jij-decompositon-original}. The susceptibilities in real space are thus given by 
\begin{equation}
    \begin{split}
        \chi_{ij} &= \frac{1}{N} \left(\sum_{k=0}^{M-1}\lambda_{ik} \chi_k \lambda_{jk} + Y \sum_{k=M}^{N-1} \lambda_{ik} \lambda_{jk} \right) \\
    &= \frac{1}{N} \left[\sum_{k=0}^{M-1}\lambda_{ik} \chi_k \lambda_{jk} + Y \left(N \delta_{ij} -\sum_{k=0}^{M-1} \lambda_{ik} \lambda_{jk} \right) \right] \,.
    \end{split}
\end{equation}
leading to Eq. \eqref{eq:chianal}.

\bibliography{main.bib}

\end{document}